# Conduction Channel Formation and Dissolution Due to Oxygen Thermophoresis/Diffusion in Hafnium Oxide Memristors

Suhas Kumar,[1]* Ziwen Wang,[2]* Xiaopeng Huang,[1]* Niru Kumari,[1] Noraica Davila,[1] John Paul Strachan,[1] David Vine,[3] A. L. David Kilcoyne,[3] Yoshio Nishi[2] and R. Stanley Williams[1]

*Equal contribution
[1]Hewlett Packard Labs, 1501 Page Mill Rd, Palo Alto, CA 94304, USA
[2]Stanford University, Stanford, CA 94305, USA
[3]Lawrence Berkeley National Laboratory, Berkeley, CA 94720, USA
Address correspondence to Suhas.Kumar@hpe.com, Stan.Williams@hpe.com

**ABSTRACT** Transition metal oxide memristors, or resistive random-access memory (RRAM) switches, are under intense development for storage-class memory because of their favorable operating power, endurance, speed, and density. Their commercial deployment critically depends on predictive compact models based on understanding nanoscale physico-chemical forces, which remains elusive and controversial owing to the difficulties in directly observing atomic motions during resistive switching, Here, using scanning transmission synchrotron x-ray spectromicroscopy to study *in-situ* switching of hafnium oxide memristors, we directly observed the formation of a localized oxygen-deficiency-derived conductive channel surrounded by a low-conductivity ring of excess oxygen. Subsequent thermal annealing homogenized the segregated oxygen, resetting the cells towards their as-grown resistance state. We show that the formation and dissolution of the conduction channel are successfully modeled by radial thermophoresis and Fick diffusion of oxygen atoms driven by Joule heating. This confirmation and quantification of two opposing nanoscale radial forces that affect bipolar memristor switching are important components for any future physics-based compact model for the electronic switching of these devices.

The recent surge in technological and commercial interest in transition-metal-oxide memristors, especially those utilizing hafnium oxide as the switching material, is accompanied by urgent efforts to formulate a compact predictive model of their behavior in large-scale integrated circuits.[1-9] Several efforts in this direction include first-principles and analytical modeling,[8,10,11] materials characterization,[2,12,13] and circuit characterization and modeling.[14,15] The resultant models are incomplete and controversial owing to a lack of understanding of the nanoscale physico-chemical forces that determine atomic motions during switching, particularly with regard to the presence and sign of temperature-gradient-driven thermophoresis of oxygen atoms, and quantification of the concentration-gradient-driven Fick diffusion.[7,8,11,16,17] Direct *in-situ* and *in-operando* studies of localized atomic motion during memristor switching can resolve these issues and improve our modeling, but such observations face steep experimental challenges due to the extremely high resolution and sensitivity required to detect atomic motions inside a functioning cell.[2,4,18,19]

In order to non-destructively study the chemical and position changes associated with oxygen atoms during memristor operation, we utilized a synchrotron-based scanning transmission x-ray microscopy (STXM) system tuned to the O K-edge with a spatial resolution of <31 nm and a spectral resolution of ~70 meV.[20] We analyzed a prototype device that had only one oxide layer to permit an unambiguous analysis of the results. To enable x-ray transmission experiments, operational memristor cells for this study were fabricated on top of 200 nm silicon nitride films suspended over holes etched into an underlying silicon wafer. The crosspoint cells were 2 µm x 2 µm in lateral dimension while the film stack within the crosspoint was (bottom to top) Pt (15 nm)/ $HfO_2$ (5 nm)/ Hf (15 nm)/ Pt (15 nm), with the $HfO_2$ as the switching layer, Hf acting as a reactive electrode and the Pt layers as the contact electrodes (Figure 1a).[21] A $HfO_x$ film within a cell that had not been operated was characterized by the O K-edge spectrum in Figure 1b, which displays a typical $HfO_x$ absorption spectrum with the component bands ($e_g$ and $t_{2g}$) marked.[1] The slight shift of the conduction band edge of the hafnium oxide in a cell with respect to that of the as-grown material may indicate a small amount of reduction after the Hf layer was deposited.

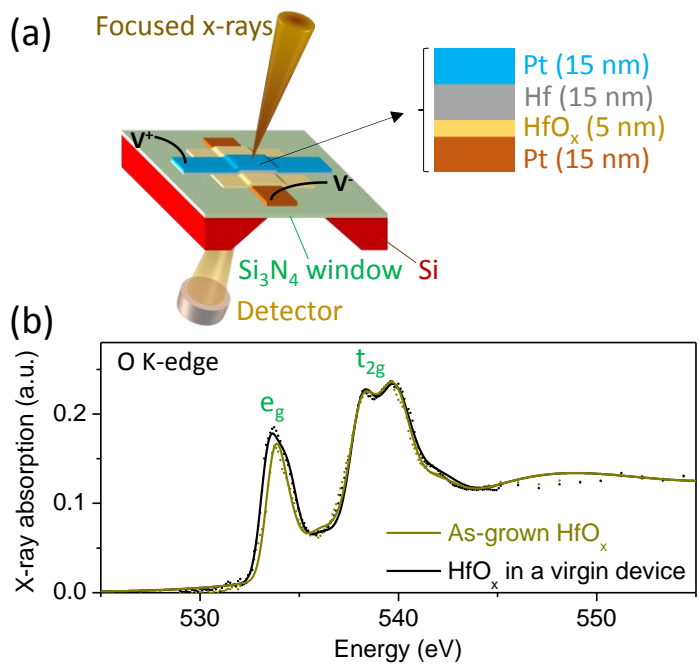

Figure 1: Schematic illustration of the experiment and oxide film spectra. (a) Drawing of the crosspoint cell geometry fabricated on top of a silicon nitride membrane. The measurement scheme using focused x-rays is also depicted. A magnified schematic cross-section of the material stack within the crosspoint cell is also shown. (b) X-ray absorption spectra of as-grown $HfO_2$ (outside a cell stack) and that of $HfO_x$ within an un-operated cell stack, in contact with the Hf reactive electrode. The main component conduction bands are identified.

## RESULTS AND DISCUSSIONS

The electrical operation of a cell (Figure 2a) involved accessing high (OFF) and low (ON) non-volatile resistance states using bipolar voltage sweeps. The first switching of the cell from its native high resistance state to a low resistance state (forming) involved a higher voltage amplitude relative to the subsequent switching behavior. Following the ON-OFF switching cycles displayed in Figure 2a, the cell was imaged utilizing several x-ray energies near the O K-edge in the STXM (570 eV post-edge image in Figure 2b) and revealed a localized region of non-uniform contrast that was absent before operation of the cell (see Supporting Information). A magnified region of the image (Figure 2c) reveals a relatively dark ring surrounding a relatively bright center. In the spectra obtained from the dark ring and the bright region (Figure 2d), the absorption intensity of the post-K-edge spectral region (>545 eV) revealed a 3-5% higher oxygen concentration in the ring and 3-5% lower oxygen concentration in the bright region, both relative to the as-grown film (see Supporting Information for details). The spectral feature at ~531 eV (below the lowest $HfO_x$ conduction band, $e_g$) within the oxygen-deficient region has been consistently assigned to oxygen vacancies, which create sub-gap defect states that provide enhanced electrical conductivity.[1,22-24] The bump in the spectrum at ~536 eV within the oxygen-rich dark region has been identified as bonds involving interstitial oxygen, previously conjectured as superoxide species.[25-27] Here, given that the as-grown $HfO_2$ lacked long-range order, we term these vacancy-like and interstitial-like defects as oxygen deficiencies and excess, respectively. For better visualization, using the relative positions of the unoccupied energy levels of the deficiencies and excess, we display schematic band diagrams of the two regions (on-ring and inside ring), that depict the conductivity enhancement due to the sub-gap defect states in the oxygen deficient region. In order to map the relative oxygen concentration across the cell, we calculated the logarithmic ratio of images obtained at a pre-O K-edge energy (522 eV) and at a post-O K-edge energy (570 eV) (Figure 2f). This plot should remove any physical distortions to the cell that may occur not related to the oxygen displacement, for example warping of the electrodes by Joule heating during the switching (which in any case was negligible for the prototype devices reported here; see Figure S17).

The radial segregation of oxygen can be explained by outward migration of oxygen atoms along the large temperature gradient created by Joule heating in an initial electronic percolation pathway created during cell 'forming', resulting in a steady-state oxygen-deficient conductive channel surrounded by a concentric oxygen excess region. Since the oxide was amorphous, it is possible that there was a small concentration of pre-existing weakly bonded oxygen that was significantly more mobile than oxygen in a single crystal (see Supporting Information), and it was this oxygen that was dislodged from the region that became deficient and transported to the ring of excess oxygen. The STXM images are not sensitive to axial oxygen displacement. The Strukov model[16] describes this process as the steady state resulting from two opposing radial forces on oxygen atoms, namely, temperature-gradient-driven thermophoresis originating from the Joule heating within the channel (directed outward from the channel to lower temperature), and concentration-gradient-driven Fick diffusion from the ring to regions with lower O concentration. The resulting oxygen segregation is 'frozen' in place upon withdrawal of current and rapid cooling to ambient temperature, thus enabling non-volatile information storage.

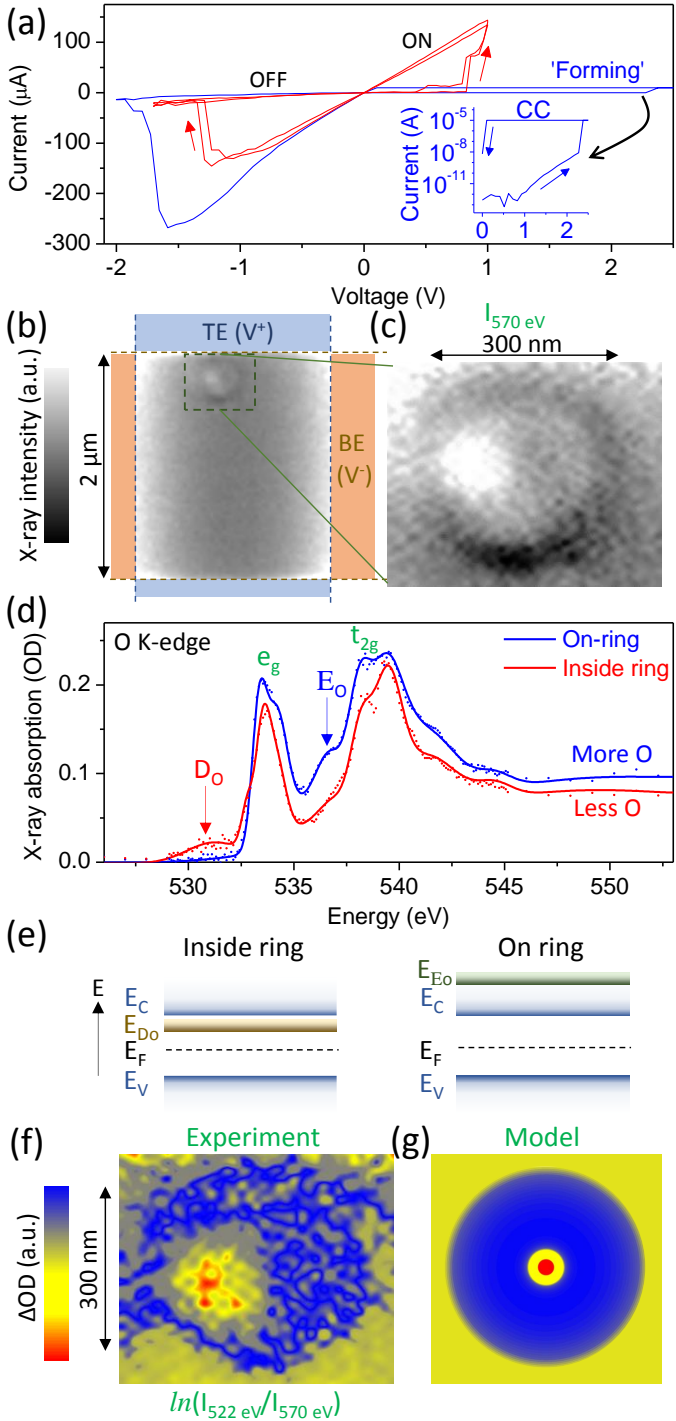

Figure 2: Cell switching and conduction channel formation. (a) Electrical switching of the resistance of the cell upon application of a voltage ramp. Blue curve labeled 'Forming' was the first switching event followed by the blue curve in the negative voltage range. The red curves show subsequent switching events between high (OFF) and low (ON) resistance states. The 'Forming' curve is shown in a logarithmic scale in the inset, with a common x-axis as the main panel. The flat curve ('CC') is the current compliance of the measurement setup. (b) X-ray transmission intensity map of the same crosspoint cell under study at an energy of 570 eV. Top electrode (TE) and bottom electrode (BE) on which the voltages were applied are marked. The region with a ring-like feature is shown in a separate magnified x-ray map in (c). (d) O K-edge spectra of the dark ring-like feature seen in (c) and of the bright central region within the feature are both substantially different from the spectra in Figure 1(b). $D_O$ corresponds to a sub-band indicating the presence of oxygen deficiencies within the central bright region and $E_O$ corresponds to a sub-band indicating the presence of oxygen excess on the dark ring-like feature. The O concentration difference between the two regions is also apparent in the post-absorption-edge region (>545 eV). (e) Schematic band diagrams of the two regions (on ring and inside ring) following the data in (d). $E_V$, $E_C$ and $E_F$ refer to energy levels of the valence band, conduction band and Fermi level, respectively, while $E_{Do}$ and $E_{Eo}$ refer to the energy levels of unoccupied antibonding orbitals originating from oxygen deficiencies and excess, respectively. The presence of $E_{Eo}$ well above $E_C$ explains why excess oxygen does not enhance electrical conductivity, whereas the sub-band-gap states of $E_{Do}$ contribute to enhanced electrical conduction. (f) Logarithmic ratio of x-ray maps obtained at 522 eV and 570 eV, representing variations in oxygen concentration (red color is lowest concentration). (g) Simulated steady-state oxygen concentration map resulting from the competition between thermophoresis and Fick diffusion.

To model the experimental results, we solved the continuity equation $\partial n_D/\partial t = \nabla \cdot J_{Fick} + \nabla \cdot J_{thermophoresis}$, where $n_D$ is the oxygen deficiency concentration, and $J_{Fick}$ and $J_{thermophoresis}$ are atomic fluxes due to Fick diffusion and thermophoresis, respectively, using a calculated steady-state temperature profile produced by Joule heating.[16,19] Both $J_{Fick}$ and $J_{thermophoresis}$ are proportional to the diffusivity $D=D_0 e^{(-E_a/kT)}$, where $k$ is the Boltzmann constant and $T$ is the absolute temperature. We chose compatible activation energies $E_a$ and diffusion constants $D_0$ from a set of non-unique values that can reasonably reproduce the ring formation in a given time (details in Supporting Information). The resultant steady state concentration profile (Figure 2g) is in fair agreement with experimental data (Figure 2f), which displays both distortion and fluctuation compared to the ideal simulation. This observation reaffirms the presence of radial thermophoresis and its direction (sign), both of which have been debated recently[7,8,11,17,19] following the original theoretical prediction of the effect.[16] This observation is also consistent with prior literature on topographical changes in TiO$_x$ memristors, where electroforming always left a round crater with a pillar in the center, the pillar being most likely the initial conducting/heating channel, from which radial temperature and O concentrations are formed.[28] In order to completely understand the physics of the switching behavior, a fully 3-dimensional temporal simulation with coupled electric-field, thermal and diffusion effects[8,11] is required, although this is not necessary for an acceptable predictive compact model sufficient for electronic circuit analysis.

In order to determine if the steady-state segregation of oxygen could be reversed by accelerating Fick diffusion at elevated temperatures (in the absence of an applied voltage), we annealed the same memristor at an ambient temperature of 200 °C for 75 minutes in nitrogen gas. STXM maps of the annealed cell (Figures 3a-3b) obtained using identical mapping parameters as those in Figure 2b-2c do not display the dark ring with the bright center seen before. The spectral signatures of the regions where the oxygen deficiency and excess regions were previously located show only minor differences from each other and are both similar to the as-grown film spectrum (Figure 1b). The dark elliptical contrast in Figures 3a-3b was confirmed to be carbon deposition on the electrodes due to the prior x-ray exposure, which did not affect the oxide film within the cell. Experimental measurements showed the oxygen concentration in this region to be uniform, while a simulation of Fick diffusion using the experimental annealing conditions reproduced the complete homogenization of oxygen segregation (Figure S9). Following the thermal annealing, the cell was electrically operated again, during which it exhibited a relatively high-voltage 'forming' step followed by ON-OFF switching, both similar to those exhibited by the cell before annealing. This further demonstrates that the homogenization of oxygen concentration drove the cell back to its native state before operation, and that the spatial separation of the oxygen deficient and excess regions in the oxide is a thermodynamically metastable state.

We further studied the temporal evolution of the electrical behavior of a cell during annealing, for which we 'formed' a fresh cell identical to the one studied using STXM, and annealed it under similar conditions as used before. The resistance (measured at room temperature) gradually increased with annealing time and saturated at about $10^{13}$ Ω (Figure 4). To estimate $E_a$, we defined a time for relaxation, $t_f$, as the time required for the resistance of a 'formed' cell (<10 kΩ) to exceed 200 GΩ upon annealing, and we assumed that the diffusivity $D \propto 1/t_f$. A plot of $ln(1/t_f)$ against $1/T$ for different temperatures yielded a straight line slope with $E_a$=0.71±0.02 eV, consistent with a theoretically predicted $E_a$ for dissolution of a conductive channel.[21] The diffusion parameters for formation and dissolution of a channel may be different (see Supporting Information for a discussion).[19,29,30]

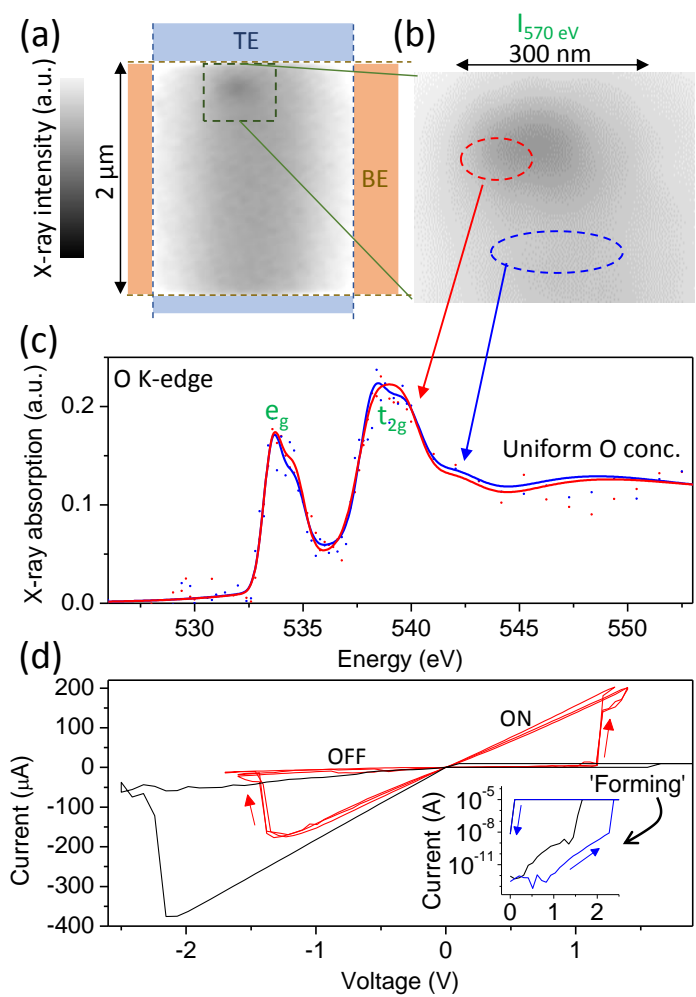

Figure 3: Conduction channel dissolution and electrical behavior. (a-b) X-ray transmission intensity map of the same crosspoint cell under study at an energy of 570 eV after annealing at 200 °C for 75 minutes. The dark region was spectrally confirmed to be carbon deposited on top of the electrode from the earlier x-ray exposure in a moderate vacuum environment. (c) O K-edge spectra of the previous locations of the ring-like feature and the center of the ring. There are no significant differences between the spectra and they are very similar to those of as-deposited films in Figure 1(b). (d) Electrical behavior of the cell, measured and plotted in an identical way as in Figure 2a. Black curves are the first two switching events with the 'Forming' curve being the first of them. The 'Forming' curve is plotted on a logarithmic scale in the inset, with a common x-axis as the main panel. The 'Forming' curve from Figure 2a is reproduced (in blue) for comparison of the previous state of the cell.

## CONCLUSIONS

In TaO$_x$ memristors, we have recently observed conduction channels similar in size and oxygen composition to those observed here using similar characterization techniques.[19]

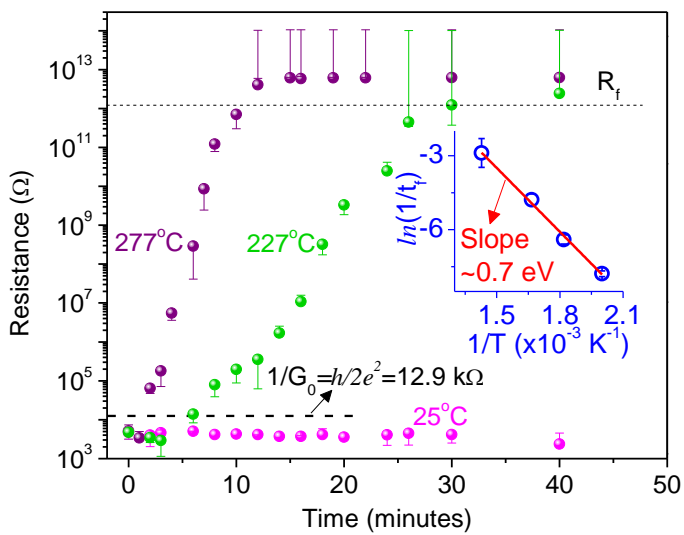

**Figure 4: Temporal evolution of resistance during annealing.** Evolution of the cell resistance for several different cells annealed at three different temperatures for about an hour measured at ~0.5 V. $G_0$ refers to the quantum conductance. $R_f$ is the defined resistance threshold for a formed cell to return to its native state. Inset is a plot of $ln(1/t_f)$ against $1/T$, whose slope corresponds to $E_a$=0.71±0.02 eV.

Although the $TaO_x$ cells were similar in construction and behavior to our $HfO_x$ cells, achieving similar local oxygen segregation required much higher power/energy levels, i.e. ~$10^5$ 5 V pulses (~7 mW) for $TaO_x$ compared to a single <2.5 V voltage sweep (<1 mW) in $HfO_x$. In addition, low power (~1 mW) switching of $TaO_x$ memristors led to uniform material changes instead of a localized channel formation.[18] There is also evidence in the literature that the activation energy for oxygen diffusion in $TaO_x$ is higher than that in $HfO_x$.[31] This highlights that oxygen migration is significantly different between these two switching oxides, which is an important consideration for material selection and predictive modeling of memristive systems. This also is possibly the origin of a trade-off between retention time and switching power between the $TaO_x$ and $HfO_x$.[32]

In conclusion, using STXM to examine operating $HfO_x$ memristor cells, we directly observed conduction channel formation after electroforming and switching by localized radial oxygen migration, and successfully modeled the phenomenon by the competition between thermophoresis and Fick diffusion. After thermal annealing, we reset the resistance of the cell back to its native state and observed that the conduction channel and accompanying ring of excess oxygen had disappeared. We re-affirmed the presence and the direction/sign of thermophoresis as a critical feature during transition-metal-oxide memristor operation and temperature as a necessary state parameter or variable in any future compact models.

## METHODS

$HfO_2$ was grown using Tetrakis(dimethylamido)hafnium (TDMA-Hf) and water as precursors at 200 °C, using a Cambridge Nanotech Fiji ALD system. The pulse time for TDMA-Hf and water were 0.25 s and 0.06 s, respectively, and $N_2$ purge time between pulses was 15 s. 50 cycles of ALD were performed and ellipsometry measurement indicated $HfO_2$ thickness of 5 nm. Bottom electrode Pt was evaporated, reactive electrode Hf was sputtered deposited and top electrode Pt was sputter deposited. Electrodes were defined by photolithography. The as-deposited film was found to be stoichiometric $HfO_2$ that was mostly amorphous. After depositing a reactive top electrode (Hf), it is very likely that the $HfO_2$ was slightly reduced at the expense of oxidation of the Hf electrode.

*Acknowledgement.* We gratefully acknowledge R. Waser and R. Dittmann for providing a detailed critique and useful suggestions to help us improve the manuscript. All synchrotron measurements were performed at the Advanced Light Source, beamlines 5.3.2.2 and 11.0.2, at Lawrence Berkeley National Laboratory, Berkeley, CA, USA. The Advanced Light Source is supported by the Director, Office of Science, Office of Basic Energy Sciences, of the U.S. Department of Energy under Contract No. DE-AC02-05CH11231. Work was performed in part at the Stanford Nanofabrication Facility which is supported by National Science Foundation through the NNIN under Grant ECS-9731293.

*Supporting Information available.* Supporting Information consists of growth of $HfO_2$, spectral processing, calculation of oxygen concentration, simulation of oxygen concentration profiles, resistance model, diffusion parameters for forming and dissolution of the filament and annealing process. This supporting information is available free of charge via the Internet at http://pubs.acs.org.

## REFERENCES AND NOTES

1. Murdoch, B. J.; McCulloch, D. G.; Ganesan, R.; McKenzie, D. R.; Bilek, M. M. M.; Partridge, J. G., Memristor and Selector Devices Fabricated from HfO2−n. *Appl. Phys. Lett.* **2016,** *108,* 143504.
2. Yang, Y.; Lu, W. D., Progress in the Characterizations and Understanding of Conducting Filaments in Resistive Switching Devices. *IEEE Trans. Nanotechnol.* **2016,** *15,* 465-472.
3. Zhou, J.; Cai, F.; Wang, Q.; Chen, B.; Gaba, S.; Lu, W., Very Low Programming-Current RRAM with Self-Rectifying Characteristics. *IEEE Electron Device Lett.* **2016,** *37,* 404-407.
4. Wedig, A.; Luebben, M.; Cho, D.-Y.; Moors, M.; Skaja, K.; Rana, V.; Hasegawa, T.; Adepalli, K. K.; Yildiz, B.; Waser, R., Nanoscale Cation

Motion in TaOx, HfOx and TiOx Memristive Systems. *Nat. Nanotechnol.* **2016,** *11*, 67-74.
5. Chen, P.-H.; Chang, K.-C.; Chang, T.-C.; Tsai, T.-M.; Pan, C.-H.; Chu, T.-J.; Chen, M.-C.; Huang, H.-C.; Lo, I.; Zheng, J.-C., Bulk Oxygen-Ion Storage in Indium-Tin-Oxide Electrode for Improved Performance of HfO2-Based Resistive Random Access Memory. *IEEE Electron Device Lett.* **2016,** *87*, 280-283.
6. Xia, L.; Gu, P.; Li, B.; Tang, T.; Yin, X.; Huangfu, W.; Yu, S.; Cao, Y.; Wang, Y.; Yang, H., Technological Exploration of RRAM Crossbar Array for Matrix-Vector Multiplication. *J. Comp. Sci. Technol.* **2016,** *31*, 3-19.
7. Gao, X.; Mamaluy, D.; Mickel, P. R.; Marinella, M., Three-Dimensional Fully-Coupled Electrical and Thermal Transport Model of Dynamic Switching in Oxide Memristors. *ECS Transactions* **2015,** *69*, 183-193.
8. Kim, S.; Choi, S.; Lu, W., Comprehensive physical model of dynamic resistive switching in an oxide memristor. *ACS Nano* **2014,** *8*, 2369-2376.
9. Wu, H.; Liao, Y.; Gao, B.; Jana, D.; Qian, H., RRAM Cross-Point Arrays. In *3D Flash Memories*, Springer: 2016; pp 223-260.
10. Zhao, L.; Clima, S.; Magyari-Köpe, B.; Jurczak, M.; Nishi, Y., *Ab Initio* Modeling of Oxygen-Vacancy Formation in Doped-Hfox RRAM: Effects of Oxide Phases, Stoichiometry, and Dopant Concentrations. *Appl. Phys. Lett.* **2015,** *107*, 013504.
11. Kim, S.; Kim, S.-J.; Kim, K. M.; Lee, S. R.; Chang, M.; Cho, E.; Kim, Y.-B.; Kim, C. J.; Chung, U. I.; Yoo, I.-K., Physical Electro-Thermal Model of Resistive Switching in Bi-Layered Resistance-Change Memory. *Sci. Rep.* **2013,** *3*.
12. Akbari, M.; Lee, J.-S., Control of Resistive Switching Behaviors of Solution-Processed HfOx-Based Resistive Switching Memory Devices by n-Type Doping. *RSC Adv.* **2016,** *6*, 21917-21921.
13. Kumar, S.; Strachan, J. P.; Kilcoyne, A. L. D.; Tyliszczak, T.; Pickett, M. D.; Santori, C.; Gibson, G.; Williams, R. S., The Phase Transition in VO2 Probed Using X-Ray, Visible and Infrared Radiations. *Appl. Phys. Lett.* **2016,** *108*, 073102.
14. Puglisi, F. M.; Qafa, A.; Pavan, P., Temperature Impact on the Reset Operation in HfO 2 RRAM. *IEEE Electron Device Lett.* **2015,** *36*, 244-246.
15. Chen, P.-Y.; Yu, S., Compact Modeling of RRAM Devices and Its Applications in 1T1R and 1S1R Array Design. *IEEE Trans. Electron Devices* **2015,** *62*, 4022-4028.
16. Strukov, D. B.; Alibart, F.; Williams, R. S., Thermophoresis/Diffusion as a Plausible Mechanism for Unipolar Resistive Switching in Metal–Oxide–Metal Memristors. *Appl. Phys. A* **2012,** *107*, 509-518.
17. Chen, C. Y.; Fantini, A.; Goux, L.; Degraeve, R.; Clima, S.; Redolfi, A.; Groeseneken, G.; Jurczak, M. In *Programming-Conditions Solutions Towards Suppression of Retention Tails of Scaled Oxide-Based RRAM*, 2015; IEEE Int. Electron Devices Meet.: pp 10-6.
18. Kumar, S.; Graves, C. E.; Strachan, J. P.; Kilcoyne, A. L. D.; Tyliszczak, T.; Nishi, Y.; Williams, R. S., In-Operando Synchronous Time-Multiplexed O K-Edge X-Ray Absorption Spectromicroscopy of Functioning Tantalum Oxide Memristors. *J. Appl. Phys.* **2015,** *118*, 034502.
19. Kumar, S.; Graves, C. E.; Strachan, J. P.; Grafals, E. M.; Kilcoyne, A. L. D.; Tyliszczak, T.; Weker, J. N.; Nishi, Y.; Williams, R. S., Direct Observation of Localized Radial Oxygen Migration in Functioning Tantalum Oxide Memristors. *Adv. Mater.* **2016,** *28*, 2772-2776.
20. Kilcoyne, A. L. D.; Tyliszczak, T.; Steele, W. F.; Fakra, S.; Hitchcock, P.; Franck, K.; Anderson, E.; Harteneck, B.; Rightor, E. G.; Mitchell, G. E.; Hitchcock, A. P.; Yang, L.; Warwick, T.; Ade, H., Interferometer-Controlled Scanning Transmission X-Ray Microscopes at the Advanced Light Source. *J. Synchrotron Radiat.* **2003,** *10*, 125-136.
21. Kalantarian, A.; Bersuker, G.; Butcher, B.; Gilmer, D. C.; Privitera, S.; Lombardo, S.; Geer, R.; Nishi, Y.; Kirsch, P.; Jammy, R. In *Microscopic Model for the Kinetics of the Reset Process in HfO2 RRAM*, 2013; IEEE VLSI-TSA Int. Symp. VLSI Technol.: pp 1-2.
22. Takeuchi, H.; Ha, D.; King, T.-J., Observation of Bulk HfO2 Defects by Spectroscopic Ellipsometry. *J. Vac. Sci. Technol. A* **2004,** *22*, 1337-1341.
23. Ramo, D. M.; Gavartin, J. L.; Shluger, A. L.; Bersuker, G., Spectroscopic Properties Of Oxygen Vacancies in Monoclinic Hf O 2 Calculated With Periodic And Embedded Cluster Density Functional Theory. *Phys. Rev. B* **2007,** *75*, 205336.
24. Chung, K. B.; Seo, H.; Long, J. P.; Lucovsky, G., Suppression of Defect States in HfSiON Gate Dielectric Films on n-Type Ge (100) Substrates. *Appl. Phys. Lett.* **2008,** *93*, 182903.
25. Ruckman, M. W.; Chen, J.; Qiu, S. L.; Kuiper, P.; Strongin, M.; Dunlap, B. I., Interpreting the Near Edges of O2 and O2- in Alkali-Aetal Superoxides. *Phys. Rev. Lett.* **1991,** *67*, 2533.
26. Jan, J. C.; Babu, P. D.; Tsai, H. M.; Pao, C. W.; Chiou, J. W.; Ray, S. C.; Kumar, K. P. K.; Pong, W. F.; Tsai, M. H.; Jong, C. A., Bonding Properties and Their Relation To Residual Stress And Refractive Index of Amorphous Ta (N, O) Films Investigated by X-Ray Absorption Spectroscopy. *Appl. Phys. Lett.* **2005,** *86*, 161910.
27. Knop-Gericke, A.; Hävecker, M.; Schedel-Niedrig, T.; Schlögl, R., High-Pressure Low-Energy XAS: a New Tool for Probing Reacting Surfaces of Heterogeneous Catalysts. *Top. Catal.* **2000,** *10*, 187-198.
28. Yang, J. J.; Miao, F.; Pickett, M. D.; Ohlberg, D. A. A.; Stewart, D. R.; Lau, C. N.; Williams, R. S., The Mechanism of Electroforming of Metal Oxide Memristive Switches. *Nanotechnol.* **2009,** *20*, 215201.
29. Kamiya, K.; Yang, M. Y.; Nagata, T.; Park, S.-G.; Magyari-Köpe, B.; Chikyow, T.; Yamada, K.; Niwa, M.; Nishi, Y.; Shiraishi, K., Generalized Mechanism of the Resistance Switching in Binary-Oxide-Based Resistive Random-Access Memories. *Phys. Rev. B* **2013,** *87*, 155201.
30. Kamiya, K.; Yang, M. Y.; Park, S.-G.; Magyari-Köpe, B.; Nishi, Y.; Niwa, M.; Shiraishi, K., ON-OFF Switching Mechanism of Resistive–Random–Access–Memories Based on the Formation and Disruption of Oxygen Vacancy Conducting Channels. *Appl. Phys. Lett.* **2012,** *100*, 073502.
31. Lee, S.; Lee, D.; Woo, J.; Cha, E.; Park, J.; Song, J.; Moon, K.; Koo, Y.; Attari, B.; Tamanna, N., Highly Reliable Resistive Switching Without an Initial Forming Operation by Defect Engineering. *IEEE Electron Device Lett.* **2013,** *34*, 1515-1517.
32. Azzaz, M.; Vianello, E.; Sklenard, B.; Blaise, P.; Roule, A.; Sabbione, C.; Bernasconi, S.; Charpin, C.; Cagli, C.; Jalaguier, E. In *Endurance/Retention Trade Off in HfOx and TaOx Based RRAM*, 2016; IEEE Int. Memory Workshop: pp 1-4.